\documentclass[aps,twocolumn,bibnotes]{revtex4-2}

\usepackage{amsmath}
\usepackage{amssymb}
\usepackage{mathtools}
\usepackage{graphicx}
\usepackage{tabularx}
\usepackage{bm}
\usepackage{amsthm}
\usepackage[toc,page]{appendix}
\usepackage[caption=false]{subfig}
\usepackage{multirow}
\usepackage{enumitem}
\usepackage{hyperref}
\usepackage{xcolor}
\usepackage{ragged2e}
\usepackage[normalem]{ulem}

\newcommand{\curl}[1]{\nabla\times #1}

\makeatletter
\renewcommand{\maketag@@@}[1]{\hbox{\m@th\normalsize\normalfont#1}}%
\makeatother

\begin{document}

\title{A Computational Picture of Hydride Formation and Dissipation In Nb SRF Cavities}

\author{Aiden V. Harbick}
\email{aharbick@byu.edu}
\author{Mark K. Transtrum}
\email{mktranstrum@byu.edu}
\affiliation{Department of Physics and Astronomy, Brigham Young University, Provo, Utah 84602, USA}

\author{Nathan S. Sitaraman}
\email{nss87@cornell.edu}
\author{Tomas A. Arias}
\author{Matthias U. Liepe}
\affiliation{Department of Physics and Astronomy, Cornell University, Ithaca, New York 14853, USA}

\date{September 16, 2025}

\begin{abstract}
   Research linking surface hydrides to Q-disease, and the subsequent development of methods to eliminate surface hydrides, is one of the great successes of SRF cavity R\&D. We use time-dependent Ginzburg-Landau to extend the theory of hydride dissipation to sub-surface hydrides. Just as surface hydrides cause Q-disease behavior, we show that sub-surface hydrides cause high-field Q-slope (HFQS) behavior. We find that the abrupt onset of HFQS is due to a transition from a vortex-free state to a vortex-penetration state. We show that controlling hydride size and depth through impurity doping can eliminate HFQS. 
\end{abstract}

\maketitle
	
\section{INTRODUCTION}
For around half a century, it has been understood that the presence of excess hydrogen  in niobium superconducting radiofrequency (SRF) cavities could lead to a variety of deleterious effects on the performance of these cavities \cite{Isagawa1980HydrogenSRF, Isagawa1980LowTNb,jisrawi1998reversible}. Much of the early work on this topic focused on methods to understand and eliminate the so-called ``Q-disease," a phenomenon caused by large surface hydrides in which the quality factor of the SRF cavity would be significantly degraded, even at low fields \cite{knobloch2003q-disease,Barkov2012DirectHydrideObservation}. More recent work on Nb-H systems has focused on understanding how hydrides contribute to high-field Q-slope (HFQS), another phenomenon in which the quality factor of a cavity becomes degraded, but only at higher fields \cite{Barkov2013HydridePrecipitation, Romanenko2013ProximityBreakdown, Romanenko2013NbSurfaceResistance}. There are two primary treatments for HFQS in Nb SRF cavities: nitrogen doping and low-temperature bakes. These methods have been studied extensively \cite{Ford2013, Grassellino2013NitrogenDoping, Checchin2020,Dhakal2020NitrogenDoping, Veit2020NitrogenDoping,Wenskat2020LowTempBake,Spina2021NitrogenDoping,Fang2023NitrogenDoping,Tamashevich2025LowTempBake}, yet the exact mechanisms for how they inhibit hydride formation, and how the hydrides themselves lead to HFQS are still open questions.

Hydride formation occurs at cryogenic temperatures in a process analogous to familiar water vapor condensation, where the high-entropy ``gas'' of interstitial hydrogen minimizes its free energy by organizing into ``droplets,'' i.e. hydride crystals. These crystals can accurately be described as low-energy ordered configurations of interstitial hydrogen with some accompanying distortion of the niobium lattice~\cite{hauck1977ordering}. In general, the physics of droplet formation is not trivial because there is a surface energy associated with the droplets which competes with the volume energy associated with the bulk phase transition. The volume energy grows with the cube of hydride radius while the surface energy grows with the square of hydride radius. Thus, for given conditions of hydrogen chemical potential and temperature, there is a ``critical" droplet radius above which hydride crystals are stable and below which they are unstable~\cite{langer1969statistical}. The fact that sub-critical droplets are unstable means that the hydrogen atoms must form a super-critical droplet purely by statistical chance, so that there is a free energy barrier to hydride precipitation which is potentially much larger than the thermal energy scale. The rate of droplet nucleation depends exponentially on this ratio and so can potentially be many orders of magnitude slower than the hopping rate of impurities.

The free energy barrier to hydride precipitation depends on the size of the critical droplet, which generally varies throughout a macroscopic sample. Of particular interest are the places where the critical droplet size and corresponding free energy barrier are small enough that hydrides can form quickly relative to the typical timescale (minutes) of cavity cooldown---we will call these places ``nucleation sites." Many material defects can potentially affect critical droplet size, including interstitial impurities, as well as more complex defects, such as impurity-vacancy complexes, dislocations and grain boundaries, that we will not describe in detail here. Impurities are of particular interest because their near-surface concentrations can be altered through low-temperature baking, and because first-principles calculations have previously shown that they create low-energy trap sites for hydrogen, potentially encouraging hydride nucleation~\cite{Ford2013}.

We present a new theory for the important physical effects of low-temperature bakes, how they improve cavity performance, and what can be done to further improve high-field quality factors. We use time-dependent Ginzburg-Landau theory to calculate dissipation from sub-surface hydrides, finding good qualitative agreement with experimentally-observed HFQS behavior~\cite{Checchin2020} and a clear relationship between hydride size, position, and HFQS onset field. We argue that increasing the concentration of hydride nucleation sites by impurity doping effectively decreases the typical size of hydrides, delaying the onset of HFQS and improving cavity performance. Our results lend additional credibility to the idea that low-temperature bakes and nitrogen doping affect high-field cavity behavior by controlling the size and distribution of hydride precipitates.

\section{Methods}\label{Methods}

\subsection{The Time-Dependent Ginzburg-Landau Equations}\label{TDGL}

Ginzburg-Landau (GL) theory is one of the oldest theories of superconductivity, and it remains relevant today owing to its relative simplicity and direct physical insights into the electrodynamic response of superconductors under static applied fields and currents \cite{GL_ref}. The \textit{time-dependent} Ginzburg-Landau (TDGL) equations were originally proposed by Schmid \cite{Schmid_TDGL} in 1966 and Gor'kov and Eliashberg \cite{GorKov_TDGL} derived them rigorously from BCS theory later in 1968. The TDGL equations (in Gaussian units) are given by:
\begin{widetext}
\begin{align}
    \Gamma\left( \frac{\partial \psi}{\partial t} + \frac{i e_s \phi}{\hbar}\psi \right) + \frac{1}{2m_s}\left(-i\hbar\nabla - \frac{e_s}{c}\mathbf{A}\right)^2\psi + \alpha \psi + \beta |\psi|^2\psi = 0 \label{psi_eq_units} \\
    \frac{4\pi \sigma_n}{c}\left( \frac{1}{c}\frac{\partial \mathbf{A}}{\partial t} + \nabla\phi \right) + \curl{\curl{\mathbf{A}}} - \frac{2\pi i e_s \hbar}{m_s c}\left( \psi^*\nabla\psi - \psi\nabla\psi^*\right) + \frac{4\pi e^2_s}{m_s c^2}|\psi|^2\mathbf{A} = 0. \label{j_eq_units}
\end{align}
\end{widetext}
These equations are solved for the complex superconducting order parameter, $\psi$, and the magnetic vector potential, $\mathbf{A}$. The magnitude squared of $\psi$ is proportional to the density of superconducting electrons. The parameters $\alpha$ and $\beta$ are phenomenological, and were originally introduced as coefficients of the series expansion of the Ginzburg-Landau free energy. Additionally, $\phi$ is the scalar potential; $\sigma_n$ is the normal electron conductivity; $\Gamma$ is the phenomenological relaxation rate of $\psi$. Furthermore, $e_s = 2e$ and $m_s = 2m_e$ represent the total charge and total effective mass of a Cooper pair, respectively. The TDGL equations are subject to boundary conditions
\begin{align}
    \left( i\hbar\nabla\psi + \frac{e_s}{c}\mathbf{A}\psi \right)\cdot \mathbf{n} = 0 \label{current_bc} \\ 
    \left(\curl{\mathbf{A}}\right)\times \mathbf{n} = \mathbf{H}_a\times \mathbf{n} \label{H_bc} \\ 
    \left( \nabla\phi + \frac{1}{c}\frac{\partial \mathbf{A}}{\partial t} \right)\cdot \mathbf{n} = 0, \label{E_bc}
\end{align}
where $\mathbf{n}$ is the outward normal vector to the boundary surface and $\mathbf{H}_a$ is the applied magnetic field.

The TDGL equations can also be derived from microscopic theory using the time-dependent Gor'Kov equations \cite{GorKov_TDGL}. A useful consequence of this derivation is that it allows the TDGL parameters to be directly related to experimentally observable properties of the material in question. The material dependencies are given by Ref. \citenum{kopnin2001theory}:
\begin{align}
    \alpha(\nu(0),T_c,T) &= -\nu(0)\left(1-\frac{T}{T_c}\right)\label{alpha_eq} \\
    \beta(\nu(0),T_c,T) &= \frac{7\zeta(3)\nu(0)}{8\pi^2T_c^2} \label{beta_eq} \\
    \Gamma(\nu(0),T_c) &= \frac{\nu(0)\pi\hbar}{8 T_c} \label{gamma_eq} \\
    m_s &= \frac{12\hbar T_c}{\pi \nu(0)v_f \ell},\label{m_eff_eq}
\end{align}
where $\nu(0)$ is the density of states at the Fermi-level, $T_c$ is the critical temperature, $T$ is the temperature, $\zeta(x)$ is the Riemann zeta function, $v_f$ is the Fermi velocity, and $\ell$ is the electron mean free path. Equation \ref{m_eff_eq} gives the effective mass specifically under the dirty limit. In order to incorporate spatially varying effective mass in the Ginzburg-Landau free energy variations, Equation \ref{psi_eq_units} should be augmented with an additional term:
\begin{equation*}
    \frac{i\hbar}{2} \nabla\frac{1}{m_s}\cdot(i\hbar \nabla + \frac{e}{c}\mathbf{A})\psi
\end{equation*}
Additionally, when solving the TDGL equations numerically, it is standard to normalize all the parameters of the model in order to obtain dimensionless quantities. To do so we introduce the following transformations:
\begin{align}
    \alpha &\longrightarrow \alpha_0 a(\mathbf{r}) \label{alpha_scale} \\
    \beta &\longrightarrow \beta_0 b(\mathbf{r}) \label{beta_scale} \\
    \Gamma &\longrightarrow \Gamma_0 \gamma(\mathbf{r}) \label{gamma_scale} \\
    m_s &\longrightarrow m_0 \mu(\mathbf{r}) \label{m_eff_scale} \\
    \sigma_n &\longrightarrow \sigma_{n0} s(\mathbf{r}). \label{sigma_scale}
\end{align}
substituting these values into Eqs. \ref{psi_eq_units} and \ref{j_eq_units}:
\begin{widetext}
\begin{align}
    \gamma\left(\frac{\partial \psi}{\partial t} + i \kappa_0\phi\psi\right) + \frac{1}{\mu}\left(\frac{-i}{\kappa_0}\nabla - \mathbf{A}\right)^2\psi + \frac{1}{\kappa_0}\nabla\frac{1}{\mu}\cdot\left(\frac{1}{\kappa_0}\nabla - i\mathbf{A}\right)\psi - a\psi + b|\psi|^2\psi = 0 \label{psi_eq}
    \\
    \frac{s}{u_0}\left(\frac{\partial \mathbf{A}}{\partial t} + \nabla\phi\right) + \curl{\curl{\mathbf{A}}} + \frac{i}{2\kappa_0 \mu}\left( \psi^* \nabla\psi - \psi\nabla\psi^*\right) + \frac{1}{\mu}|\psi|^2\mathbf{A} = 0, \label{current_eq}
\end{align}
\end{widetext}
where $\kappa_0 = \frac{\lambda_0}{\xi_0}$ is the Ginzburg-Landau parameter of the reference material. The quantity $\lambda_0 = \sqrt{\frac{m_0 c^2 \beta_0}{4\pi e^2_s |\alpha_0|}}$ is the penetration depth of the reference material, and $\xi_0 = \sqrt{\frac{\hbar^2}{2m_0|\alpha_0|}}$ is its coherence length. The parameter $u_0 = \frac{\tau_{\psi_0}}{\tau_{j_0}}$ is a ratio of characteristic time scales in the reference material, where $\tau_{\psi_0} = \frac{\Gamma_0}{|\alpha_0|}$ is the characteristic relaxation time of $\psi$ in the reference material and $\tau_{j_0} = \frac{\sigma_{n0}m_0\beta_0}{e_s^2|\alpha_0|}$ is the characteristic relaxation time of the current. We have also inserted a minus in front of $a$, which is just a convention to make positive values of $a$ correspond to the superconducting state (Note: this is the opposite of how $\alpha$ is usually interpreted within Ginzburg-Landau theory, however it is standard to make this change when performing nondimensionalization). The boundary conditions become:
\begin{align}
    \left( \frac{i}{\kappa_0}\nabla\psi + \mathbf{A}\psi \right)\cdot n = 0 \\
    \left(\curl{\mathbf{A}}\right)\times n = \mathbf{H}_a\times n
    \\
    \left( \nabla\phi + \frac{\partial \mathbf{A}}{\partial t} \right)\cdot n = 0.
\end{align}

\subsection{Calculating Values for the TDGL Parameters via DFT}

To determine appropriate values for $a(\mathbf{r})$, $b(\mathbf{r})$, $\gamma(\mathbf{r})$, and $\mu(\mathbf{r})$, we will consider experimental results from the literature and perform \textit{ab initio} density-functional theory (DFT) calculations where direct experimental measurements are unavailable. From Eqs. \ref{alpha_eq}-\ref{m_eff_eq}, we know these parameters mostly depend on well-defined microscopic quantities, namely $\nu(0)$, $T_c$, $v_f$, and $\ell$ all of which can be calculated using DFT. To do so we use the JDFTx software package with the PBE exchange-correlation functional and ultrasoft pseudopotentials \cite{Sundararaman,PBE,Psp}. In order to perform precise integrals over the Fermi surface, we use the Wannier function method as implemented in the FeynWann package for JDFTX \cite{Wannier,Wannier2,sundararaman2014theoretical,brown2016ab}. Fermi-surface integrals immediately give values for $\nu(0)$ and $v_f$, while $T_c$ is calculated by calculating the phonon dispersion, applying Eliashberg theory, and using the McMillan formula with $\mu * = 0.12$ \cite{eliashberg1960interactions,mcmillan1968transition}. To estimate electron mean free path $\ell$, we consider electron scattering off of the perturbing potential of a hydrogen vacancy as the likely principal defect in NbH, with a defect fraction of $0.075$ corresponding to the hydrogen-poor compositional limit of the $\beta$ hydride phase \cite{ricker2010evaluation}.

Niobium hydride calculations used an 8-atom fcc unit cell with $6 \times 6 \times 9$ k-point folding. Fermi surface integrals for $\nu(0)$, $v_f$, and $\ell$ used an energy-conserving delta function width of 5mH, and used the Wannier method to interpolate the k-space mesh to a 16x finer k-space mesh. The Fermi surface integral for $T_c$ used an energy-conserving delta function width of 0.74mH and a 24x finer k-space mesh. For perturbing potentials of phonon modes and hydrogen vacancy defects, we used a 96-atom supercell of this unit cell. a planewave cutoff energy of 20 Hartree, and an effective electron temperature of 5mH.

Experiment can then give information about the compositions of sample materials, and DFT calculations can determine the $\nu(0)$, $v_f$, and $T_c$ associated with these compositions. Using these values in addition to estimates of the electron mean free path (which can be derived from DFT or can come from experimental characterizations), $a(\mathbf{r})$, $b(\mathbf{r})$, $\gamma(\mathbf{r})$,and $\mu(\mathbf{r})$ are calculated from Eqs. \ref{alpha_eq}-\ref{m_eff_eq}, and the material geometries from the experimental results determine the spatial variation.

\subsection{Dissipation in TDGL}

When simulating SRF materials, dissipation is often a physical quantity of interest. While TDGL allows us to estimate dissipation, it is important to emphasize that both the dissipation calculations and the quality factor estimates that follow lie well outside the regime of quantitative validity for the theory. In particular, TDGL is strictly valid only near $T_c$ in the gapless limit, and its predictions for dissipation under RF-like dynamic fields at low temperatures should be interpreted with caution. Despite this, we believe the calculations presented here remain qualitatively valuable. They provide a means of linking mesoscopic-scale simulations to macroscopic cavity performance metrics, and enable relative comparisons between different material configurations that may inform experimental priorities. 

Under TDGL, a formula for dissipation can be derived by considering the time derivative of the free energy and the free energy current flux density. A more detailed derivation is found in Ref. \citenum{kopnin2001theory}, but we quote the final result here:
\begin{equation}
    D = 2\Gamma\left|\left(\frac{\partial \psi}{\partial t} + \frac{ie_s\phi\psi}{\hbar}\right)\right|^2 + \sigma_n \mathbf{E}^2. \label{dissipation_eq}
\end{equation}
This quantity is a power density, with the first term corresponding to the superconducting dissipation arising from the relaxation of the order parameter. The second term is the dissipation of normal currents which are largest near the surface where magnetic field can still appreciably penetrate.

A particularly relevant quantity that can be estimated from the dissipation is the cavity quality factor, $Q$, which is given by
\begin{equation}
    Q = \frac{2\pi E}{\Delta E}, \label{quality_factor_eq}
\end{equation}
where $E$ is the energy stored in the cavity and $\Delta E$ is the energy dissipated in the cavity walls each RF period. It is common to express the quality factor as
\begin{equation} \label{reduced_quality_factor_eq}
    Q = \frac{G}{R_s},
\end{equation}
where $R_s$ is the cavity surface resistance and $G$ is a geometric factor that depends only on quantities which are determined by the cavity geometry. For a typical 1.3 GHz 9-cell Nb TESLA cavity, $G = 270$ $\Omega$ \cite{Nb_SRF_Cavity_Info}. The surface resistance is given by
\begin{equation}
    R_s = \frac{\mu_0 \omega \lambda^3}{\Tilde{H_a^2} L_x L_y}\left(I_\psi + \omega \frac{\sigma_n \mu_0 \lambda^2 T_{sim}}{2\pi} I_A\right), \label{surface_resistance_eq}
\end{equation}
where $\mu_0$ is the permeability of free space, $\omega$ is the cavity frequency, $\lambda$ is the penetration depth, $\Tilde{H_a}$ is the maximum applied magnetic field value in simulation units, $L_x$ and $L_y$ are the size of a simulation domain in the X and Y directions respectively, $\sigma_n$ is the normal conductivity, and $T_{sim}$ is the period of the applied field in simulation time units. $I_\psi$ and $I_A$ are integrals over the squared time derivatives of $\psi$ and $\mathbf{A}$:
\begin{align}
     I_\psi &\equiv \int d\Tilde{t} \int d\Tilde{x} \int d\Tilde{y} \int d\Tilde{z} \left|\frac{\partial\Tilde{\psi}}{\partial \Tilde{t}}\right|^2 \\
     I_A &\equiv \int d\Tilde{t} \int d\Tilde{x} \int d\Tilde{y} \int d\Tilde{z} \left(\frac{\partial\Tilde{\mathbf{A}}}{\partial \Tilde{t}}\right)^2
\end{align}
where the tilde variables denote ones which are in simulation units. A much more detailed derivation of these equations can be found in Reference \citenum{Harbick2025SampleSpecific}.

The value of Q calculated from TDGL outputs will typically be underestimated at low field. This is because of the assumption of gapless superconductivity, which results in higher surface resistances than is predicted with the BCS surface resistance. Despite this, our approach still often predicts quality factors within an order of magnitude of the experimental values. Additionally, the relative behavior of Q at different applied fields qualitatively captures effects such as high field Q-slope. Nonetheless, we emphasize that this quality factor calculation is best treated as a qualitative tool; for quantitatively accurate predictions, more rigorous superconductivity theories should be used.

\subsection{Estimating Effective TDGL Parameters} \label{estimating_effective_parameters}

In certain situations, a system may behave as if it were a single, homogeneous material, even if it consists of multiple distinct components. One such case occurs when numerous small defects, such as nano-hydrides, are distributed across a surface, leading to a collective behavior that resembles a new effective material. This subsection outlines the method we developed to extract the effective TDGL parameters of such systems, allowing us to estimate how the material's properties evolve when influenced by a large number of small, interacting defects.

The following discussion focuses on the case of a large number of nano-hydrides, though the general approach we describe can be applied to any system in which collective effects result in a new uniform effective material. To model this behavior, we simulate a system with a specific hydrogen concentration, using a smaller domain size of $2\lambda \times 2\lambda \times 8\lambda$ to keep computational costs manageable. The extended $z$-direction ensures that the order parameter, $\psi$, can fully reach its zero-field value without experiencing finite-size effects. Hydrides with a radius of $0.05\lambda$ are randomly placed within the domain, chosen from a uniform distribution, and added until the desired hydrogen concentration is reached, with no overlap between hydrides.

After placing the hydrides, we solve for the TDGL order parameter $\psi$ at two distinct field conditions: zero field (to estimate the zero-field value, $\psi_\infty$) and a low, non-zero field (to extract the coherence length, $\xi$). The zero-field value, $\psi_\infty$, provides the ratio $\alpha/\beta$ through the relation:
\begin{equation}
|\psi_\infty|^2 = \frac{\alpha}{\beta},
\end{equation}
which is given by Tinkham \cite{tinkham2004introduction}. However, this relation does not independently determine $\alpha$ or $\beta$. To resolve this, we calculate the coherence length, $\xi$, which depends on $\alpha$. By determining $\xi$ from the low-field solution of $\psi$, we can then extract the value of $\alpha$, and from there, calculate $\beta$ using the ratio $\alpha/\beta$. The coherence length, $\xi$, is related to $\alpha$ by:
\begin{equation}
\xi^2 = \frac{\hbar^2}{2m_s\alpha},
\end{equation}
again based on Tinkham \cite{tinkham2004introduction}.

To estimate $\xi$, we calculate the average value of $\psi$ in the $xy$-plane for each value of $z$ at low (but non-zero) field. For a uniform material, the order parameter would follow the equation:
\begin{equation}
\psi(z) = \psi_\infty - Ce^{-\sqrt{2}(z_{max} - z)/\xi},
\end{equation}
where $z_{max}$ is the $z$-coordinate of the top of the domain where the field is applied, with $C$ and $\xi$ as the fitting parameters. The fitted value of $\xi$ can be used to estimate the coherence length. However, since our system is not truly uniform and $\psi_\infty$ varies slightly due to the random distribution of hydrides, directly using this equation would lead to poor fits. Instead, we modify the approach by fitting the difference $\Delta \psi(z)$ between the calculated $\psi(z)$ and the local value of $\psi_\infty(z)$, where $\psi_\infty(z)$ is the zero-field value of $\psi$ averaged over the $xy$-plane for each value of $z$. This modification effectively filters out the noise and isolates the exponential decay associated with $\xi$:
\begin{equation}
\Delta\psi(z) = \psi(z) - \psi_\infty(z) = - Ce^{-\sqrt{2}(z_{max} - z)/\xi}.
\end{equation}

With $\xi$ now estimated, we can use it to determine the effective value of $\alpha$ from the relationship between $\xi$ and $\alpha$, as described earlier. From this, we can also estimate $\beta$ using the ratio $\alpha/\beta$, which was determined from the zero-field solution for $\psi_\infty$. Once the effective $\alpha$ and $\beta$ are determined, we can estimate the critical field $H_c$ using the relation:
\begin{equation}
H_c^2 = \frac{4\pi \alpha^2}{\beta}.
\end{equation}
Since $\kappa$ can be expressed as a function of $\beta$:
\begin{equation}
\kappa^2 = \frac{m_s^2 c^2 \beta}{2\pi \hbar^2 e_s^2}, \label{Hc_eq}
\end{equation}
we now have the necessary parameters to calculate the superheating field. The superheating field for the composite material is then estimated using the formula for low $\kappa$ materials \cite{Transtrum_Hsh}:
\begin{equation}
H_{sh} = 2^{-3/4} \kappa^{-1/2} \frac{1 + 4.6825120 \kappa + 3.3478315 \kappa^2}{1 + 4.0195994 \kappa + 1.0005712 \kappa^2}. \label{Hsh_eq}
\end{equation}
This formula provides an estimate of the superheating field for a system containing many small hydrides, which collectively behave as an effective material with a new $\kappa$.

\subsection{Geometry and Numerical Approach for TDGL Simulations}

\begin{figure}[htbp]
    \centerline{\includegraphics[width=0.95\columnwidth]{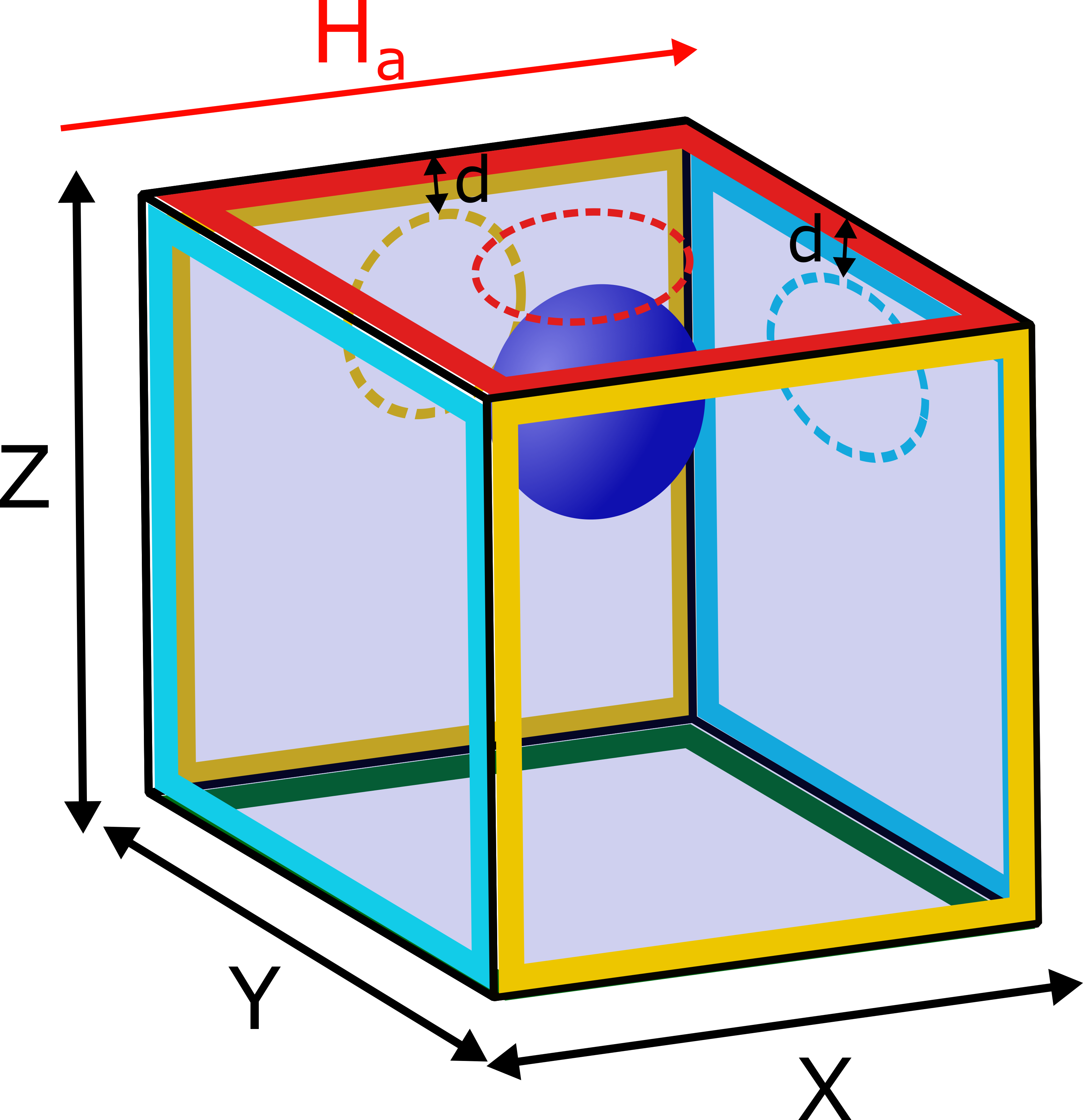}}
    \caption{\justifying Schematic of the simulation geometry (not to scale). Surfaces outlined in yellow and light blue represent the periodic boundary conditions in the $x$ and $y$ directions, respectively. An external field is applied to the surface outlined in red, in the direction indicated by the arrow labeled ``$H_a$". No field is applied to the surface outlined in green. The hydride is modeled as a dark blue sphere, with dotted lines showing the projections of the sphere onto the $xy$, $yz$, and $zx$ planes. The color of the dotted circle corresponds to the plane in which the projection occurs. The distance from the surface to the outer edge of the hydride, denoted as $d$, represents the island’s position relative to the surface.}
    \label{fig:SimulationGeom}
\end{figure}

A schematic of the geometry used in our TDGL simulations is shown in Fig. \ref{fig:SimulationGeom}. The simulation domain for the large hydride simulations is a $40\lambda \times 40\lambda \times 20\lambda$ cuboid, with periodic boundary conditions applied in the x and y directions (highlighted in yellow and light blue, respectively). An external field is applied to the upper surface in the z-direction (highlighted in red), while the bottom surface (highlighted in green) is free from any applied field. The hydrides, modeled as spheres, are depicted in the figure, with dotted circles showing the projections of the spheres onto the XY, YZ, and ZX planes. These projections help orient the reader to the spatial arrangement of the hydride in three-dimensional space. The distance from the surface to the outer edge of the hydride, denoted by $d$, is marked on the figure and represents one of the key parameters varied in our simulations to explore how the hydride’s position influences its impact on cavity performance. For the nano-hydride simulations, we used a domain of size $2\lambda \times 2\lambda \times 8\lambda$, with the same periodic boundary conditions and applied field.

The simulations were conducted using cubic meshes generated with the open-source mesh generation tool, Gmsh \cite{geuzaine2009gmsh}. The OpenCASCADE geometry kernel within Gmsh was employed to adapt the mesh to the shape of the hydride islands. To solve the TDGL equations, we applied the finite element formulation proposed by Gao \cite{gao3D}. All computations were performed using the open-source finite element software FEniCS \cite{alnaes2015fenics}. A detailed analysis of the methods used for solving the TDGL equations can be found in Harbick and Transtrum \cite{Harbick2025SampleSpecific}.

\subsection{Qualitative Random Walk Simulation of Hydride Formation} \label{Random_walk_methods}

To illustrate the possible relationship between impurity-induced disorder and hydride size, we developed a simple random walk simulation of hydride formation. In this simulation, hydride nucleation sites exist at a concentration of $10^{-2}$ at the surface, and this concentration decays exponentially to $2*10^{-6}$ far beneath the surface. We vary the exponential decay constant to simulate different impurity diffusion depths. Hydrogen atoms exist at a concentration of $5*10^{-3}$ and undergo random walk movement on the $60 \times 60 \times 1000$ cubic simulation lattice, and become frozen when they are either adjacent to a nucleation site or adjacent to a previously-frozen hydrogen atom.

While this simulation is not rooted in physical atomic interactions and operates at a scale too small to simulate hydrides of the sizes considered in our Ginzburg-Landau simulations, it adequately illustrates the inverse relationship between local nucleation site concentration and typical hydride size. This relationship is important as we will ultimately consider the implications of our Ginzburg-Landau simulation results for developing new experimental recipes. Ongoing research seeks to develop a more physically realistic model of the effect of impurity doping on hydride size distribution.

\section{RESULTS}

\subsection{Nucleation Sites and Hydride Formation}

\begin{figure}[htbp]
    \centering
    \includegraphics[width=2.5in]{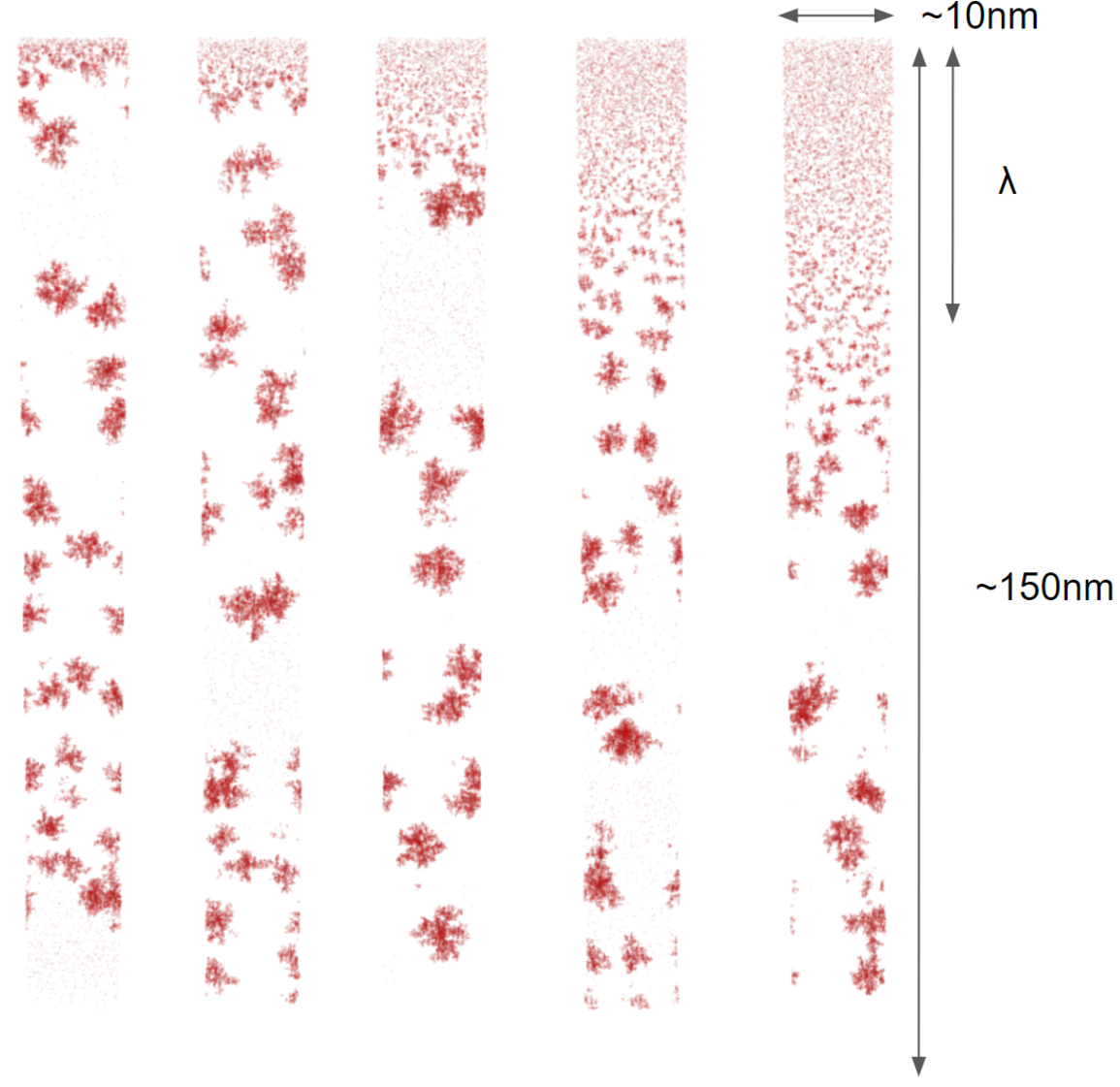}
    \caption{
        \justifying Qualitative simulations of hydride nucleation for different exponential nucleation site distributions, from shallowest (left) to deepest (right).
        }
    \label{fig:hydride_formation_sim}
\end{figure}

The distribution of nucleation sites can greatly influence the size and location of hydrides. To illustrate this, we consider a simple classical model of hydride nucleation, which begins with a uniform concentration of free hydrogen interstitial impurities and an exponential profile of nucleation sites, a description of this method is found in Section \ref{Random_walk_methods}. In this model, hydrogen atoms freeze if they arrive at a site adjacent to a nucleation site, or if they arrive at a site adjacent to a frozen hydrogen interstitial. To model a niobium surface which has been impurity-doped to some degree, we take the concentration of nucleation sites to be simply proportional to the concentration of impurities. Figure~\ref{fig:hydride_formation_sim} shows the results of this model. We find that a shallow doping depth results in large hydrides near the surface, while a deeper doping depth results in much smaller hydrides near the surface. Generally, hydride size is inversely proportional to nucleation site concentration, as expected.

These results demonstrate there are, roughly speaking, two regimes for hydride precipitation: One in which there are a small number of larger, mesoscopic-scale hydrides, and another regime in which there is a large number of microscopic scale nano-hydrides. The following two subsections will address both these regimes. 

\subsection{TDGL Simulations of Mesoscopic Hydrides}

To estimate the impact of mesoscopic-scale hydrides, we use density-functional theory to calculate material properties of hydrides~\cite{sitaraman2022theory}, and then we perform time-dependent Ginzburg-Landau (TDGL) theory simulations of hydride dissipation, the results of which are detailed in this section. The material values used in our simulations are shown in Table \ref{tab:Material_Values}.

\begin{table}[htbp]
  \centering
  \begin{tabular}{@{}lcc@{}}
    \hline\hline
    Quantity & Nb value & NbH value \\
    \hline
    $T_c$ (K)                                & 9    & 0.5 \\
    $\nu(0)$ (states/(eV\,nm$^{3}$))         & 90   & 25  \\
    $v_f$ ($10^{5}$ m/s)                     & 6.15 & 6.2 \\
    $\ell$ (nm)                              & 40   & 20  \\
    \hline\hline
  \end{tabular}
  \caption{\justifying A summary of the material values used in the TDGL simulations.
  The $T_c$, $\nu(0)$, and $v_f$ values were calculated with DFT. We assumed that the Nb mean free path is comparable to a coherence length, and the mean free path in the hydride is half of the bulk value.}
  \label{tab:Material_Values}
\end{table}

We find that hydrides have a low-field state in which they dissipate more energy per unit volume than the superconducting niobium, resulting in a lower low-field quality factor $Q_0$. This dissipation is simply the result of normal currents of Bogoliubov quasiparticles moving through a material of finite resistivity; it does not cause any noticeable Q-slope, and for realistic hydride concentrations the overall effect on dissipation is small.

\begin{figure}[htbp]
    \centering
    \includegraphics[width=2.5in]{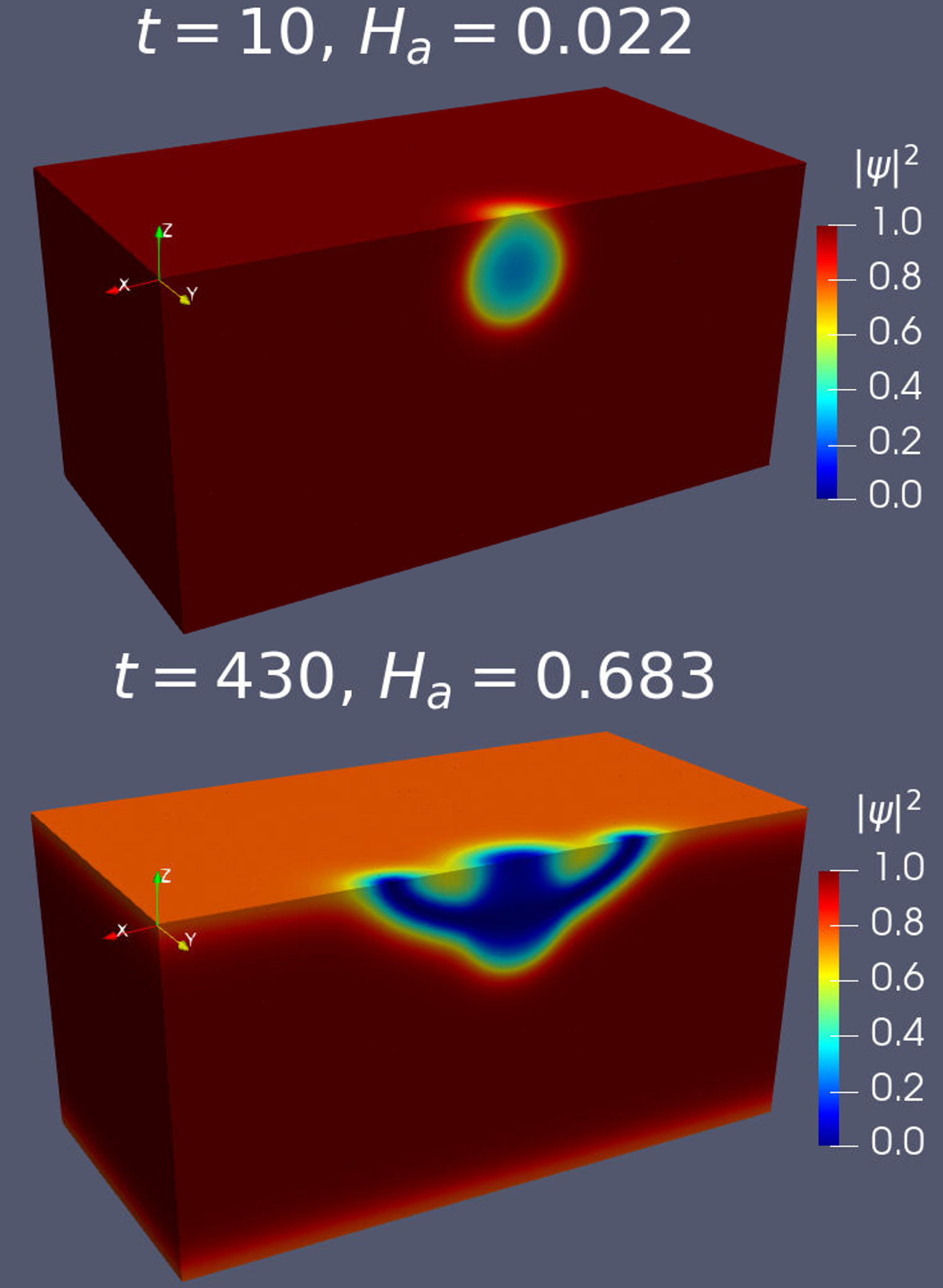}
    \caption{
        \justifying Simulations of superconducting order parameter at low field (top), showing a weak spot at the surface, and at high field (bottom) showing flux vortex entry.
        }
    \label{fig:Vortex_nucleation}
\end{figure}

We find that hydrides have a fundamentally different high-field state in which a more complicated dissipation mechanism occurs, involving penetration of flux vortices. The transition from the low-field state to the high-field state is associated with an abrupt increase in calculated energy dissipation, or an abrupt onset of Q-slope, at a critical value of the peak magnetic field. Vortex penetration occurs because the proximity-coupling effect affects the superconducting properties of the niobium surface above a sub-surface hydride. This creates a weak spot where flux vortex penetration can occur at fields significantly below the superheating field~(Fig.~\ref{fig:Vortex_nucleation}).

\begin{figure}[htbp]
    \centering
    \includegraphics[width=0.85\linewidth]{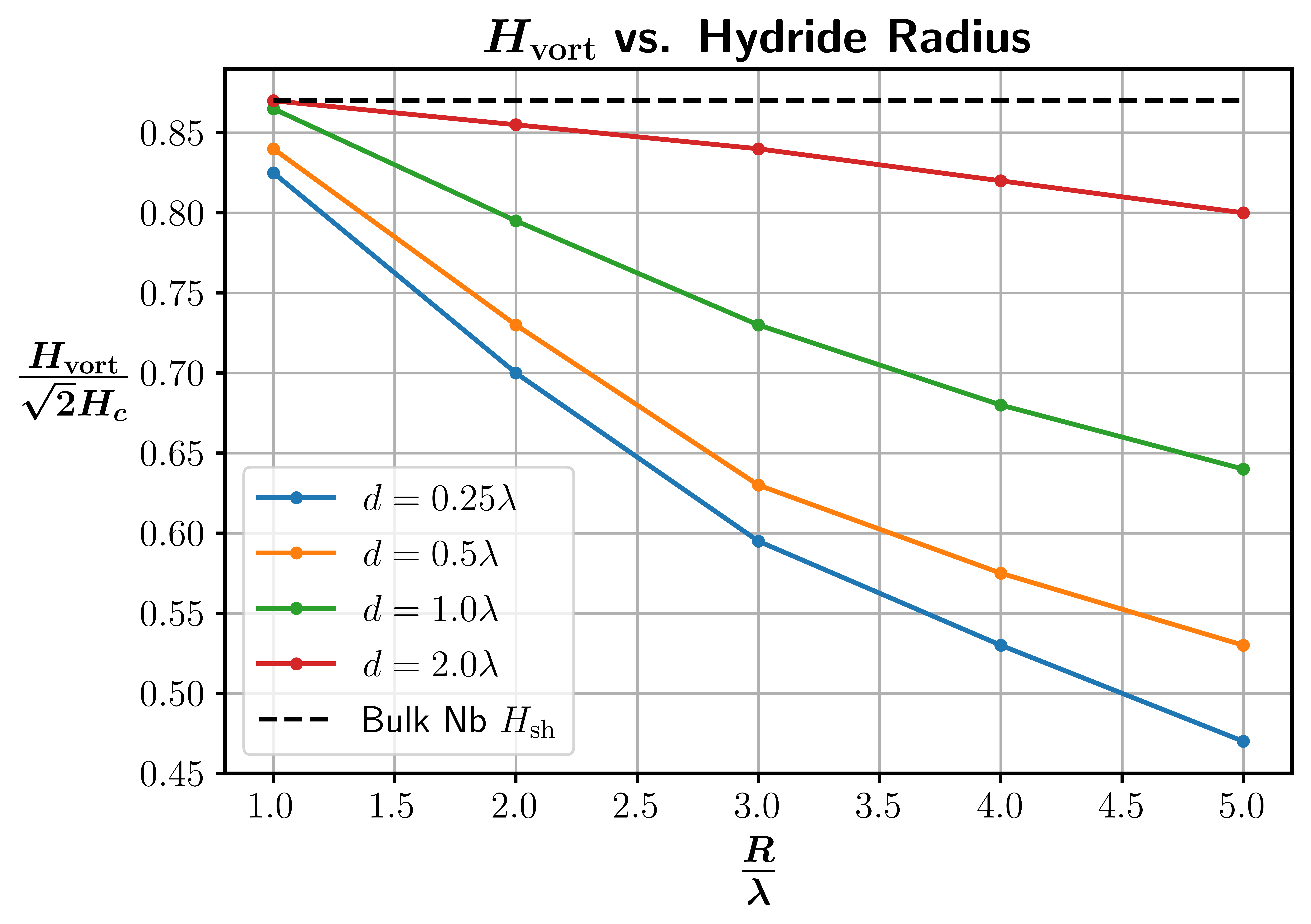}
    \caption{
        \justifying Calculated vortex entry field as a function of hydride radius for hydrides at different depths.
        }
    \label{fig:Hvort_vs_R}
\end{figure}

\begin{figure}[htbp]
    \centering
    \includegraphics[width=0.85\linewidth]{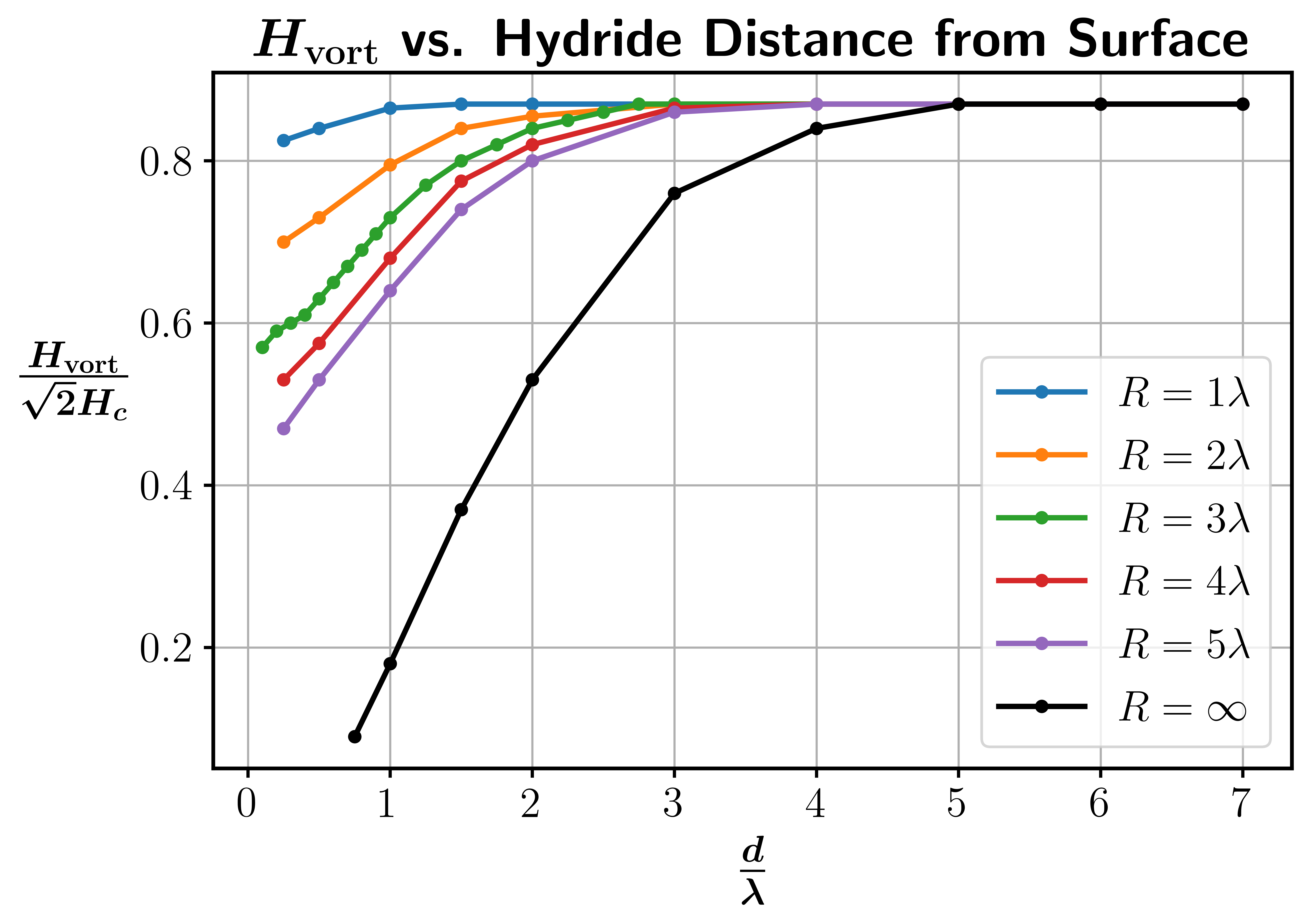}
    \caption{
        \justifying Calculated vortex entry field as a function of hydride depth for hydrides of different radii, the infinite radius hydride was simulated using a layered simulation with a layer of Nb of thickness d on top of a NbH layer.
        }
    \label{fig:Hvort_vs_d}
\end{figure}

Dissipation from hydrides in the vortex state, unlike dissipation from hydrides in the non-vortex state, is highly field-dependent. This can be explained in part by the fact that, as the field increases beyond the critical field for vortex entry into the hydride, the fraction of the RF cycle at or above this critical field increases rapidly, thus increasing the length of time per cycle that vortex-related dissipation occurs. Additionally, we find that the number of vortices entering a hydride increases with increasing field, further increasing dissipation. Together, these effects result in a steep Q-slope beyond the critical field for vortex entry.

Hydride size and proximity to the surface play a crucial role in determining the critical field for vortex entry into the hydride. Figure \ref{fig:Hvort_vs_R} shows the vortex penetration field $H_{vort}$ versus Hydride radius ($R$). The vortex penetration field consistently decreases with respect to hydride size, meaning that larger hydrides will achieve the vortex state at lower field thresholds. Additionally, we see that the decrease in $H_{vort}$ with respect to hydride radius becomes larger as the hydrides form closer to the surface. This can also be seen in Figure \ref{fig:Hvort_vs_d}, which depicts $H_{vort}$ versus $d$. Finite-size hydrides with radius $R = 5\lambda$ nucleated vortices $\sim50\%$ below the bulk superheating field value. We additionally calculated $H_{vort}$ for a hydride of infinite size. This was done by simulating a layer of Nb of thickness $d$ on top of a hydride layer which, due to the periodic boundary conditions, represents a hydride of infinite size. The lowest thickness layer simulated this way was for $d = 0.75\lambda$, which resulted in $H_{vort} = 0.09\sqrt{2}H_c$, for values of $d$ smaller than this, the Nb layer was unable to maintain superconductivity and the whole system would quench for any nonzero applied field. The infinite hydride represents a limiting case for the impact that hydrogen can have on SRF performance, and the behavior we observe in our simulations is consistent with the ``Hydrogen Q-disease" which is attributed to large hydride precipitates.

\begin{figure}[htbp]
  \centering
  \subfloat{\includegraphics[width=0.8\linewidth]{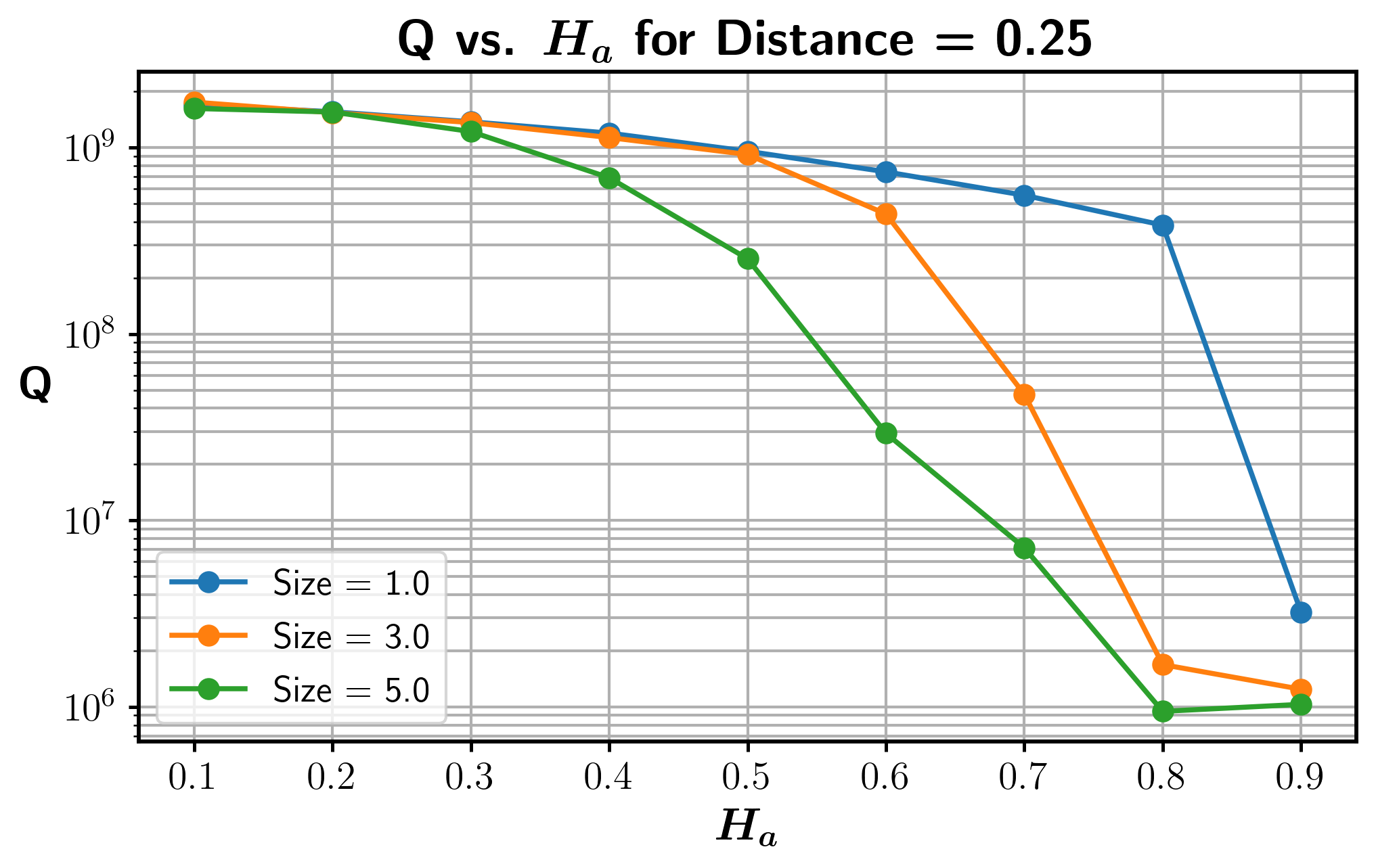}}\par\vspace{1em}
  \subfloat{\includegraphics[width=0.8\linewidth]{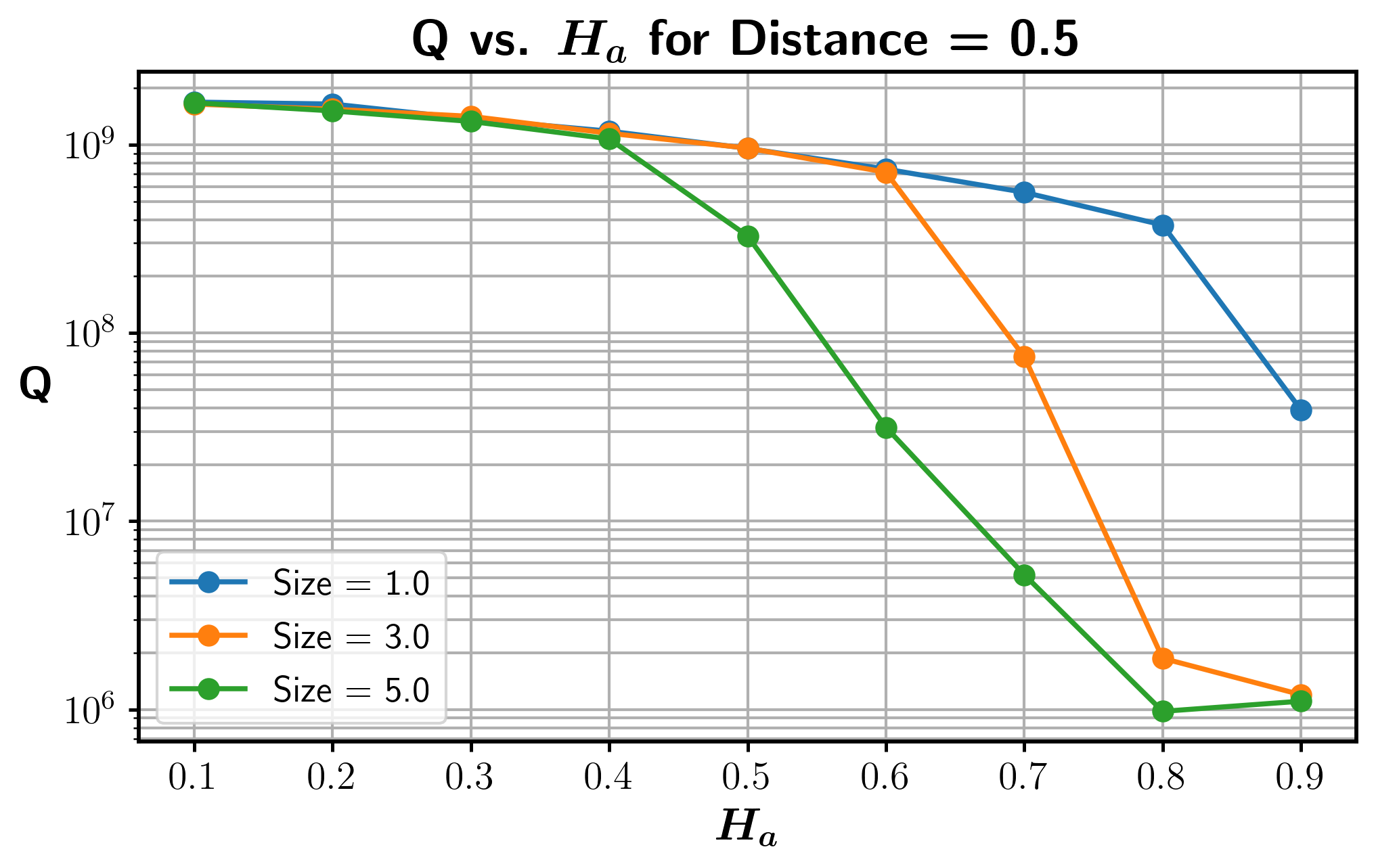}}\par\vspace{1em}
  \subfloat{\includegraphics[width=0.8\linewidth]{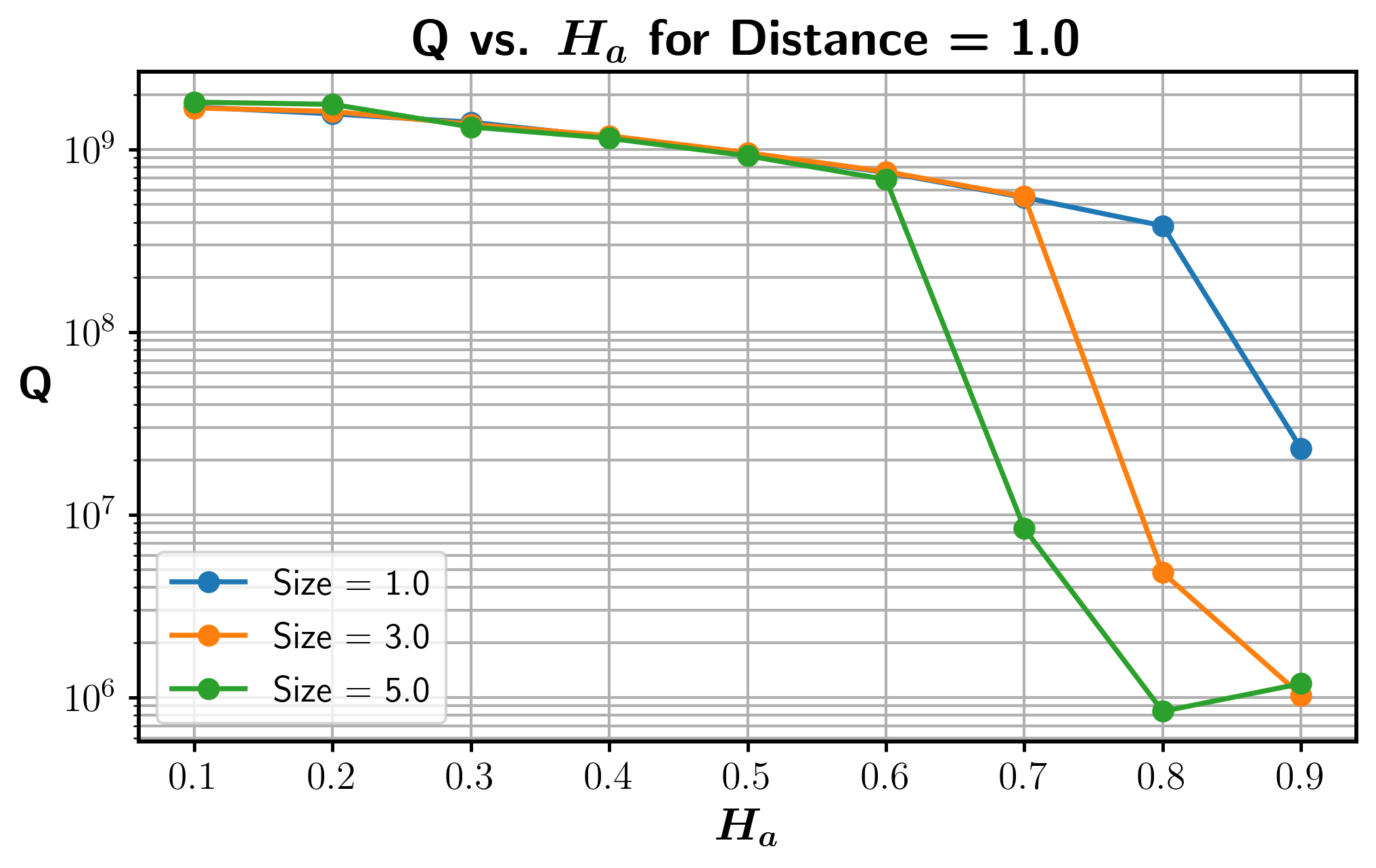}}\par\vspace{1em}
  \subfloat{\includegraphics[width=0.8\linewidth]{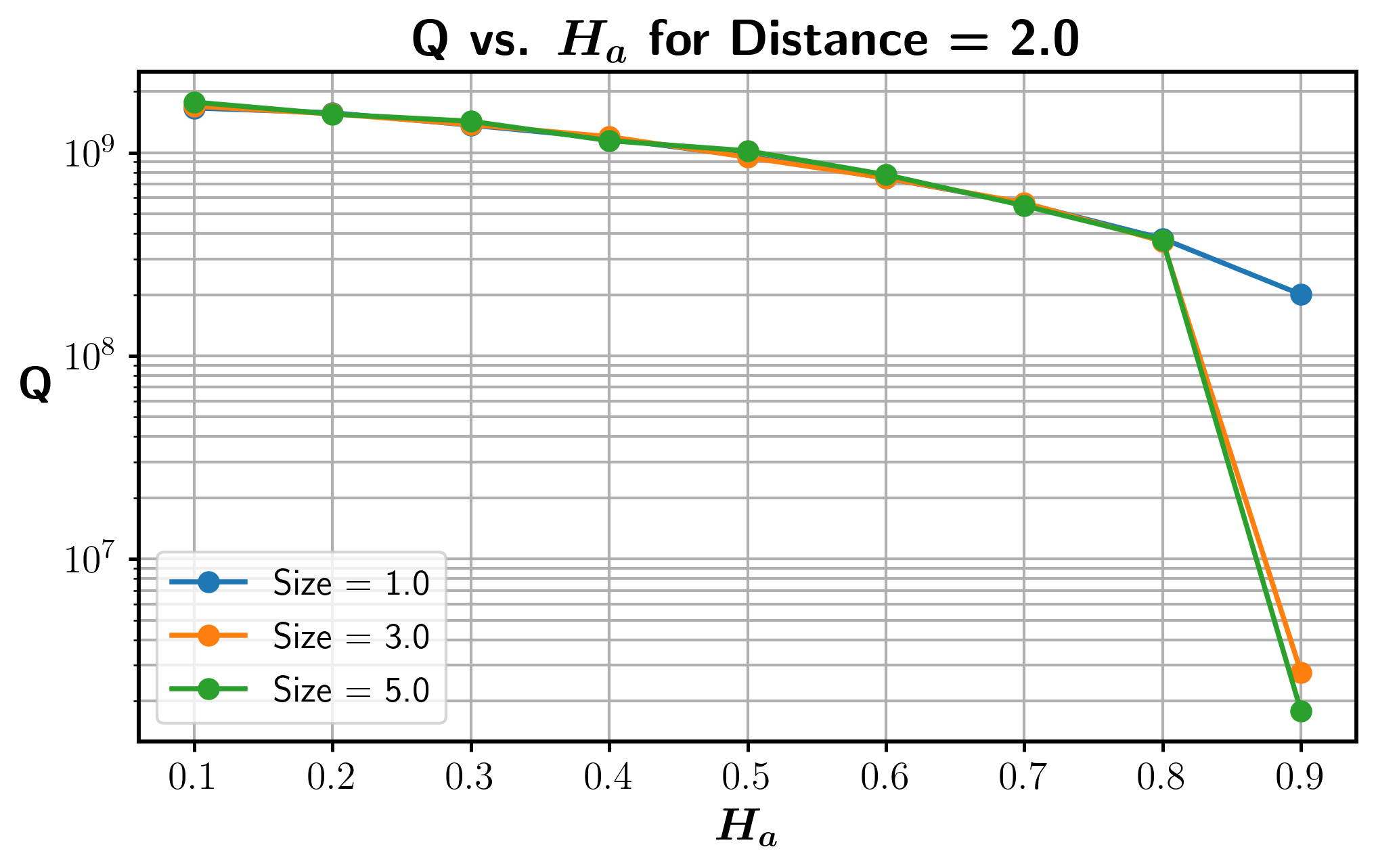}}
  \caption{\justifying Calculated quality factor as a function of field for hydrides at different depths (shallowest at top to deepest at bottom) and of different sizes (different colored lines in each plot).}
  \label{fig:Q_vs_ha}
\end{figure}

The distribution of hydrides has an important effect on cavity quality factor at high fields, specifically by altering the adverse high-field Q-slope (HFQS) behavior. Our results indicate that even modest changes to the characteristic size of near-surface hydrides can explain experimentally observed changes in the onset field of the high-field Q-slope~(Fig.~\ref{fig:Q_vs_ha})~\cite{Checchin2020}.

The $Q$-factor calculations in Fig. \ref{fig:Q_vs_ha} assume that each cavity contains exclusively one type of hydride, with fixed size and position. To relax this assumption, we can incorporate multiple types of hydrides, making certain assumptions about their distribution within the cavity. By combining the dissipation estimates for each hydride type, we can calculate a weighted average of the dissipation values according to the distribution of hydride types. This yields a new composite $Q$-slope that reflects the impact of the entire hydride distribution. 

\begin{figure}[htbp]
    \centering
    \includegraphics[width=0.85\linewidth]{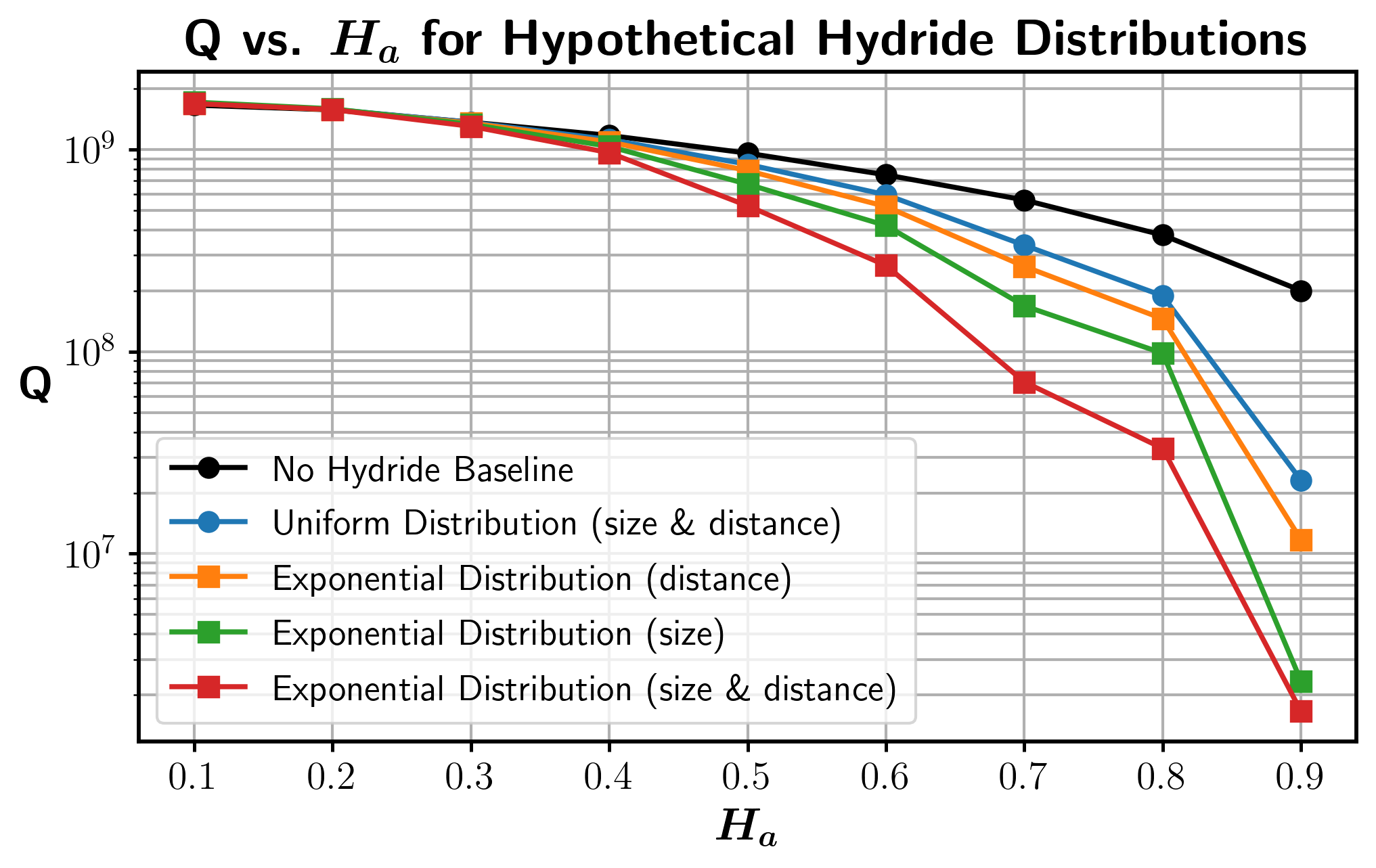}
    \caption{
        \justifying Hypothetical quality factor versus applied field curves for different hydride distributions. The black curve represents the baseline simulation with no hydrides. The blue curve assumes a uniform distribution of both hydride size and position. The yellow curve assumes a uniform hydride size distribution and an exponential distribution of hydride distances, with a higher likelihood of hydrides near the surface. The green curve assumes a uniform distribution in distance and an exponential distribution in hydride size, with larger hydrides being more probable. The red curve assumes both hydride size and distance follow exponential distributions, with large hydrides near the surface being most common.
    }
    \label{fig:Q_vs_Ha_Hypothetical}
\end{figure}

Figure \ref{fig:Q_vs_Ha_Hypothetical} shows simulations based on these distributions. The black curve represents the quality factor for a simulation with no hydrides. The blue curve corresponds to a uniform distribution of both hydride size and position, meaning all defects are equally likely. The yellow curve assumes hydride size is uniformly distributed, while the distance between hydrides follows an exponential distribution, with a higher likelihood of hydrides near the surface. The green curve assumes a uniform distance distribution, but with hydride sizes exponentially distributed, meaning larger hydrides are more common. Finally, the red curve assumes both hydride size and distance follow exponential distributions, with large hydrides near the surface being most frequent.

These distributions are idealized, and future work could incorporate more detailed experimental characterizations of hydride size and position distributions, as well as simulations of a wider variety of hydride types. Nevertheless, the current results remain qualitatively informative: the simulated quality factor curves are consistent with those observed experimentally. This supports our assertion that the dissipation mechanism responsible for the high-field $Q$-slope in some Nb cavities is hydride-induced vortex nucleation, and that eliminating large, near-surface hydrides is key to reducing high-field quality slope (HFQS) in Nb SRF cavities. As shown in Fig. \ref{fig:hydride_formation_sim}, procedures designed to reduce hydride size might instead lead to the formation of a large number of nano-hydrides. While individual nano-hydrides may have negligible effects, we now turn to explore the potential collective effects when a large number of nano-hydrides are present.

\subsection{Nano-hydride simulations} \label{nano-hydride_simulations}

In this section, we perform TDGL simulations to investigate the collective effects of a large number of nano-hydrides. The simulations were conducted using the methods described in Section \ref{estimating_effective_parameters}, with a domain size of $2\lambda \times 2\lambda \times 8\lambda$. A nano-hydride radius of $0.05\lambda$ was chosen as a representative case for hydrides that, on their own, would be considered negligible. We ran simulations varying the number of nano-hydrides from $0$ to $2000$, corresponding to hydrogen concentrations between $0\%$ and $3.2\%$. This range was chosen based on the expectation that typical local hydrogen concentrations in SRF cavities would not exceed $3\%$.

The hydrogen concentration, $C_H$, is estimated using the formula:
\begin{equation}
C_H = \frac{V_H}{V_{Nb} + V_H},
\end{equation}
where $V_H$ is the volume of the hydrides and $V_{Nb}$ is the volume of niobium in the simulation domain. In this case, the entire domain contains niobium, so $V_{Nb} = 32$, and $V_H = \frac{4}{3}\pi (0.05)^3 N$, where $N$ is the number of hydrides in the given simulation.

\begin{figure}[htbp]
  \centering
  \subfloat{%
    \includegraphics[width=0.8\linewidth]{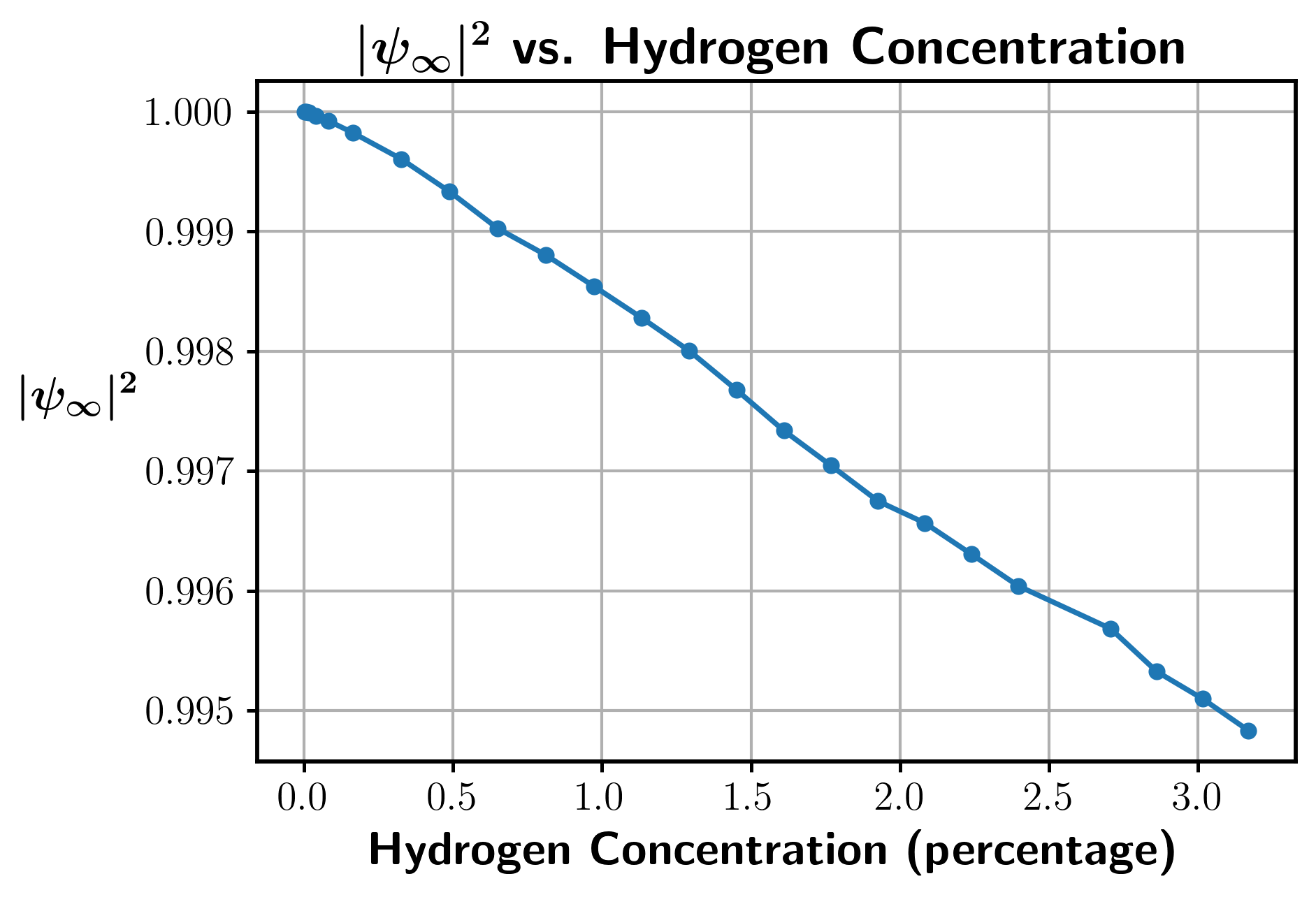}%
  }\par\vspace{1em}
  \subfloat{%
    \includegraphics[width=0.8\linewidth]{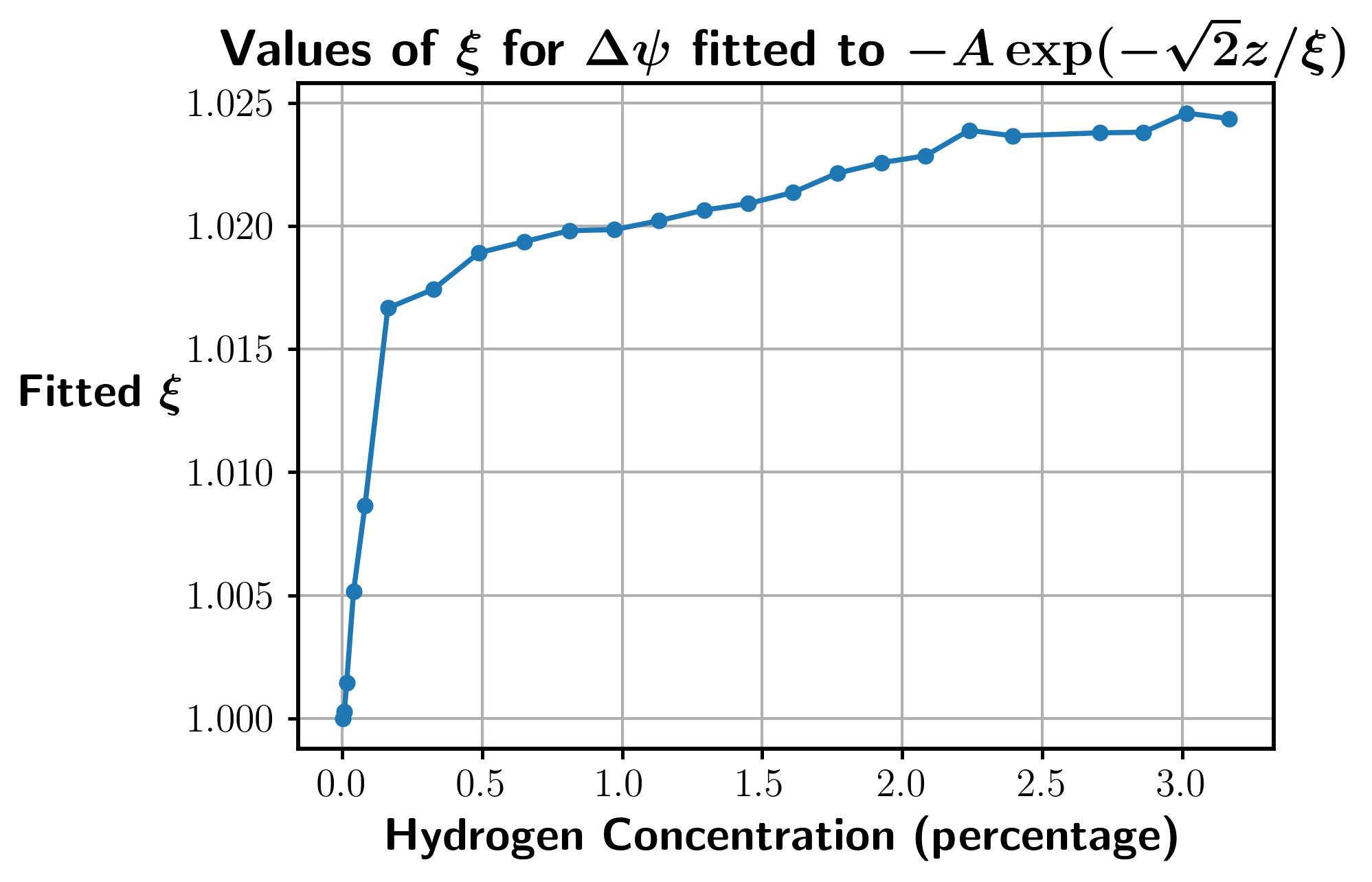}%
  }
  \caption{\justifying Calculated values of $|\psi_\infty|^2$ (top) and $\xi$ (bottom) as functions of hydrogen concentration.}
  \label{fig:psi_inf_xi}
\end{figure}

\begin{figure}[htbp]
    \centering
    \includegraphics[width=0.85\linewidth]{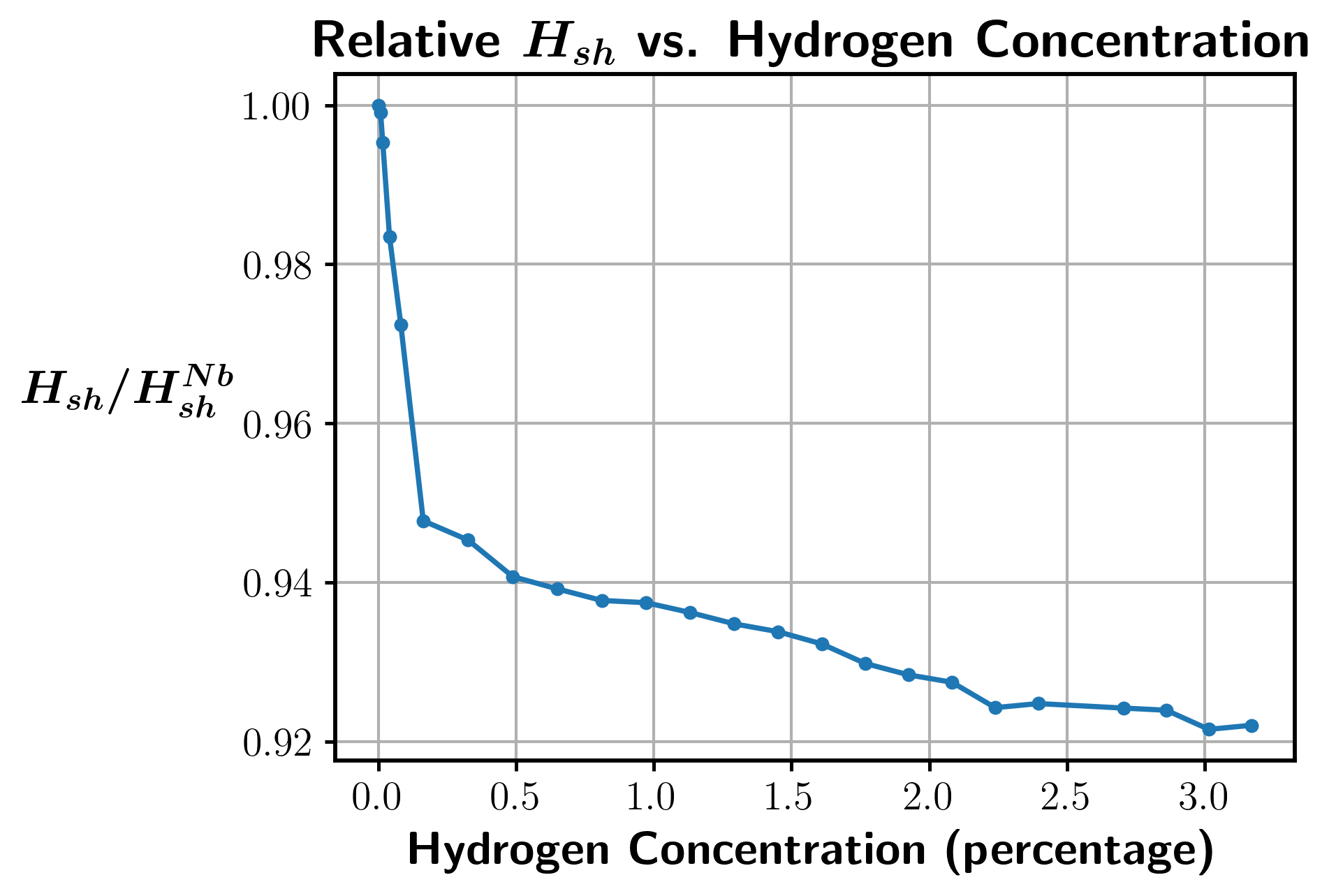}
    \caption{
        \justifying Plot of relative $H_{sh}$ versus hydrogen concentration. The relative $H_{sh}$ is shown with respect to the superheating field of pure niobium, $H_{sh}^{Nb}$. A decrease of approximately $5\%$ in $H_{sh}$ is observed once the hydrogen concentration exceeds $0.2\%$, slowly decreasing on average with increasing hydrogen concentration.
        }
    \label{fig:Relative_Hsh_vs_H_Concentration}
\end{figure}

Following the methods outlined in Section \ref{estimating_effective_parameters}, we calculate $|\psi_\infty|^2$ and $\xi$ as functions of hydrogen concentration (see Fig. \ref{fig:psi_inf_xi}). From these values, we estimate $\alpha$ and $\beta$ relative to the corresponding values for Nb, and then use Eqs. \ref{Hc_eq} and \ref{Hsh_eq} to estimate the critical field $H_c$ and superheating field $H_{sh}$. The resulting superheating field estimate is shown in Figure \ref{fig:Relative_Hsh_vs_H_Concentration}, where we plot $H_{sh}$ relative to the superheating field of pure niobium, $H_{sh}^{Nb}$.

For hydrogen concentrations below approximately $0.2\%$, $H_{sh}$ quickly decreases by around $5\%$ from $H_{sh}^{Nb}$. This is observed in simulations with fewer than $100$ hydrides, where the number of hydrides is insufficient to create an effective new material. Once the number of hydrides exceeds $100$, the hydrides begin to collectively behave as an effective medium, and $H_{sh}$ stabilizes at about $5\%$ below $H_{sh}^{Nb}$. Beyond this point, $H_{sh}$ continues to gradually decrease with increasing hydrogen concentration. The fluctuations in $H_{sh}$ are due to variability in the $\xi$ fits across different simulations.

The overall decrease in $H_{sh}$ due to the collective effects of nano-hydrides is around $5\%-7\%$. This change is significantly smaller than the impact of larger, single hydrides, supporting the idea that smaller hydrides, even in large quantities, result in improved cavity performance. This finding offers a potential explanation for how low-temperature baking and nitrogen doping procedures can reduce high-field $Q$-slope by promoting the formation of smaller near-surface hydrides, which are less detrimental to cavity performance.

\section{CONCLUSIONS}

In this study, we investigated the effects of hydride size and position on the high-field $Q$-slope (HFQS) in Nb SRF cavities. We performed TDGL simulations of hydrides with varying sizes and distributions, focusing on how these factors influence the superconducting properties of the cavities. Our results show that hydrides, particularly those near the surface, play a significant role in the onset of HFQS. We found that larger hydrides, which transition into the vortex state at lower fields, are the primary contributors to the observed dissipation at high fields. In contrast, smaller hydrides, even in large quantities, have a much smaller impact on vortex nucleation and do not significantly affect the cavity performance.

We expect that in a realistic surface with a distribution of hydride sizes and positions, there will be a gradual transition from low-field behavior with little to no $Q$-slope to high-field behavior with a steep $Q$-slope at fields where ``typical" near-surface hydrides transition into the vortex state. This trend is qualitatively consistent with experimental $Q$-slope measurements in low-temperature baked cavities, as well as measurements in ``clean" cavities, which do not undergo baking and have very low impurity content. Cavities with low impurity content or relatively short exponential doping profiles typically show a low-field state with little $Q$-slope, followed by a distinct high-field quality slope (HFQS) state. As noted by other researchers, the onset of HFQS is directly related to impurity doping \cite{Bafia2022}, and our model is consistent with this observation if the characteristic size of hydrides is inversely proportional to impurity concentration.

We emphasize that this mechanism differs subtly from previous proposals, in which impurities were thought to trap near-surface hydrogen and prevent hydride formation. While it is unlikely that impurities can completely prevent hydride formation by trapping hydrogen, our results indicate that it is not necessary to entirely eliminate hydrides to reduce HFQS. The key factor is eliminating large hydrides, which transition into the vortex state significantly below the niobium superheating field. Therefore, counterintuitively, creating more hydride nucleation sites near the surface can be beneficial by reducing the characteristic size of hydrides.

Our simulations support the idea that decreasing the size of hydrides is the most important factor for improving Nb SRF cavity performance, with the distance from the surface being the next most important factor. The dissipation mechanism behind HFQS in our simulations is hydride-induced vortex nucleation, which contrasts with the mechanism proposed in a 2013 study by Romanenko et al. \cite{Romanenko2013ProximityBreakdown}, which suggested that proximity breakdown in the hydrides was responsible for HFQS.

While caution is needed in quantitatively interpreting quality factor and dissipation from TDGL simulations, as these are primarily qualitative tools, the simulations provide valuable insights into the underlying mechanisms. The steady-state properties, such as critical fields, offer more quantitative validity, particularly in the dirty limit where our simulations are most relevant.

Overall, our results provide new insights into the role of hydrides in the dissipation mechanisms that cause HFQS in Nb SRF cavities. We hope that these findings can guide future cavity construction and performance refinement, particularly in optimizing hydride size and distribution to improve cavity performance at high fields.
	
\bibliographystyle{ieeetr}
\bibliography{main}
	
\end{document}